\documentclass[reprint, amsmath,amssymb,showpacs,floatfix, aps, prb, twocolumn, superscriptaddress
]{revtex4-2}

\usepackage{graphicx}% Include figure files
\usepackage{dcolumn}% Align table columns on decimal point
\usepackage{bm}% bold math
\usepackage{color}

\usepackage{hyperref}
\hypersetup{colorlinks=true,citecolor=blue,linkcolor=blue, urlcolor=blue}

\def\nn{\nonumber}
\def\({\left(}
\def\){\right)}
\def\[{\left[}
\def\]{\right]}
\def\a{\alpha}

\def\d{\delta}
\def\e{\epsilon}

\def\s{\sigma}
\def\w{\omega}

\def\bfQ{\mathbf{Q}}

\def\bfe{\mathbf{e}}
\def\bfk{\mathbf{k}}
\def\bfp{\mathbf{p}}
\def\bfq{\mathbf{q}}

\def\>{\rangle}
\def\<{\langle}

\begin{document}

\title{First-Principles Ultrafast Exciton Dynamics and Time-Domain Spectroscopies:\\ Dark-Exciton Mediated Valley Depolarization in Monolayer WSe$_2$
}

\author{Hsiao-Yi Chen}
\affiliation {Department of Applied Physics and Materials Science, 
California Institute of Technology, Pasadena, California 91125}
\affiliation{Department of Physics, California Institute of Technology, Pasadena, California 91125}
\affiliation{RIKEN Center for Emergent Matter Science (CEMS), Wako, Saitama, 351-0198, Japan}

\author{Marco Bernardi}
\email{bmarco@caltech.edu}
\affiliation {Department of Applied Physics and Materials Science, 
California Institute of Technology, Pasadena, California 91125}
\affiliation{Department of Physics, California Institute of Technology, Pasadena, California 91125}
%\date{}

\begin{abstract}
Calculations combining first-principles electron-phonon ($e$-ph) interactions with the Boltzmann equation enable studies of ultrafast carrier and phonon dynamics. However, in materials with weak Coulomb screening, electrons and holes form bound excitons and their scattering processes become correlated, posing additional challenges for modeling nonequilibrium physics. 
Here we show calculations of ultrafast exciton dynamics and related time-domain spectroscopies using $ab~initio$ exciton-phonon (ex-ph) interactions together with an excitonic Boltzmann equation.  
Starting from the nonequilibrium exciton populations, we develop simulations of time-domain absorption and photoemission spectra that take into account electron-hole correlations. 
We use this method to study monolayer WSe$_2$, where our calculations predict sub-picosecond timescales for exciton relaxation and valley depolarization and reveal the key role of intermediate dark excitons. 
The approach introduced in this work enables a quantitative description of nonequilibrium dynamics and ultrafast spectroscopies in \mbox{materials} with strongly bound excitons.
\end{abstract}
\maketitle
\section{Introduction}
\vspace{-10pt}
First-principles methods based on density \mbox{functional} theory (DFT)~\cite{martin2020electronic,baroni2001phonons} can characterize electron-phonon ($e$-ph) interactions, enabling quantitative studies of nonequilibrium electron dynamics~\cite{jhalani2017ultrafast,tong2021toward,caruso2021nonequilibrium} and transport properties in materials ranging from semiconductors to organic crystals and correlated electron systems~\cite{bernardi2016first,li2015electrical,zhou2016ab,lee2018charge,zhou2021ab}. 
However, these methods focus on the independent dynamics of electron and hole carriers, whereas in many semiconductors, wide-gap insulators, and nanostructured materials, where the Coulomb interaction is weakly screened,  excited electrons and holes can form charge-neutral bound states (excitons) which dominate optical response and light emission~\cite{knox1963theory,wang2018colloquium,mueller2018exciton}. 
\\
\indent
Exciton are key to many scientific and technological advances $-$ exciton diffusive dynamics governs the efficiency of energy and light-emitting devices~\cite{gregg2003excitonic,menke2014exciton}, and excitons trapped in two-dimensional (2D) materials can provide stable optical qubits~\cite{thureja2022electrically}. 
Excitons can additionally carry spin and valley quantum numbers, with potential applications to information storage and processing in spintronic and valleytronic devices~\cite{gunawan2006valley,xiao2007valley,schaibley2016valleytronics}. 
Atomically-thin transition metal dichalcogenides (TMDs) have become a widely used platform for experimental studies of exciton physics due to their robust excitonic effects persisting up to room temperature~\cite{mueller2018exciton,wang2018colloquium}.  
\\
\indent %optical and photoemission 
Time-domain spectroscopies can probe the energy and internal structure of excitons, and characterize their interactions and dynamics down to the femtosecond timescale~\cite{bertoni2016generation,madeo2020directly, cho2005exciton,wang2018colloquium}.
Yet, microscopic interpretation of these experiments is difficult, especially in cases where excitons are present. Therefore the development of theoretical methods to study exciton dynamics and its spectroscopic signatures remains a priority. 
\\
\indent
Although analytical and semiempirical models have been proposed to study exciton dynamics~\cite{toyozawa1958theory,toyozawa1964interband,segall1968phonon,rustagi2018photoemission}, predictive first-principles calculations would be desirable to interpret novel experiments in this rapidly evolving arena. 
Much first-principles work on excitons has focused on improving the description of their binding energy, optical response, and radiative lifetime, typically using the \textit{ab initio} Bethe-Salpeter equation (BSE) approach~\cite{rohlfing2000electron,onida2002electronic,spataru2004electronic,
palummo2015exciton,chen2018theory,jhalani2019precise}. 
Combined with linear-response DFT, this framework has recently enabled calculations of exciton-phonon (ex-ph) interactions and the associated phonon-induced exciton relaxation times and photoluminescence (PL) linewidths~\cite{chen2020exciton, Antonius-PRB2020}.  These advances set the stage for real-time simulations of exciton dynamics and time-domain spectroscopies.
\\
\indent
Here we show calculations of ultrafast exciton dynamics based on exciton properties and ex-ph interactions computed from first principles with the BSE and validated via the PL linewidth. 
Our real-time simulations employ a bosonic Boltzmann transport equation (BTE) to evolve in time the exciton populations and characterize phonon-induced exciton relaxation. 
We apply this method to monolayer $\rm WSe_2$, where we determine the ultrafast timescales for phonon-induced bright-to-dark exciton relaxation ($\sim$0.5 ps at 300~K) and exciton valley depolarization (185 fs at 77~K and 65 fs at 300~K). By comparing these results with hole carrier dynamics in the single-particle picture, we show that dark excitons can debottleneck intervalley exciton scattering and speed it up by orders of magnitude. %
Using the exciton populations as input, we also predict time-resolved angle-resolved photoemission (tr-ARPES) and transient absorption spectra including excitonic effects. 
This work demonstrates a quantitative framework for exciton dynamics and sheds light on ultrafast optical processes in \mbox{2D-TMDs.} 
\newpage

\section{Methods}
%\vspace{-10pt}
\subsection{Exciton band structure} 
\begin{figure}[!b]
\centering
\includegraphics[scale=0.23]{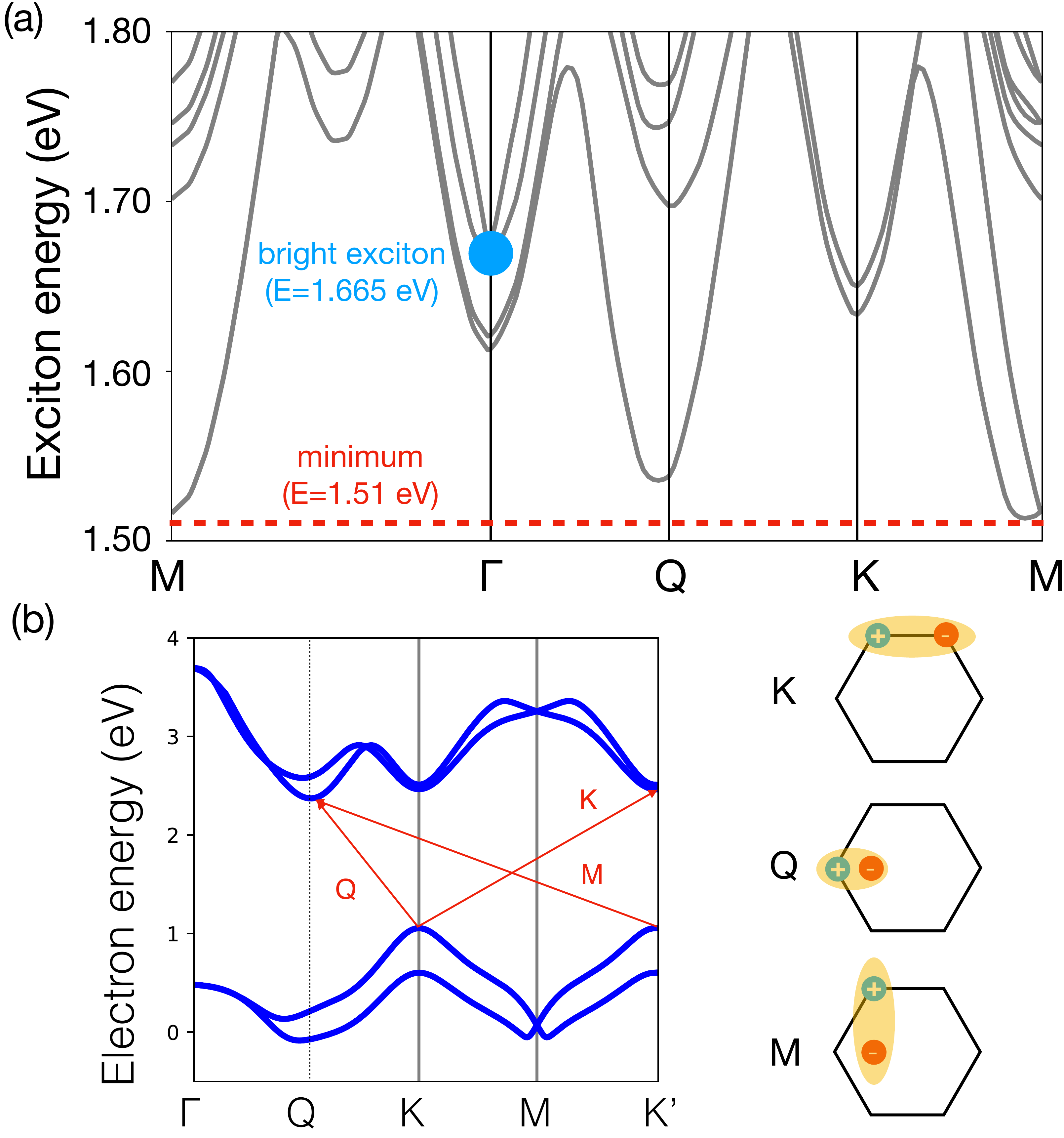}
\caption{(a) Exciton energy vs. momentum dispersion in monolayer WSe$_2$, with minima at M and Q.  
The two lowest bright excitons are the 3$^{\rm rd}$ and 4$^{\rm th}$ states at $\Gamma$, with energy $E_0=1.665$ eV indicated with a blue dot. 
(b) Electronic band structure showing the transitions that make up dark excitons with $\mathbf{Q}=(\rm{K},~\rm{Q},~\rm{M})$ (left) and the corresponding electron-hole pairs shown schematically in the electronic BZ (right).} 
\label{Fig:band}
\end{figure}
\vspace{-13pt}
The effective mass approach has been widely used to model exciton dynamics~\cite{selig2016excitonic,brem2020phonon}, but now one can compute the full exciton band structure from first principles using DFT plus the finite-momentum BSE~\cite{Qiu,Cudazzo,sponza2018direct,chen2020exciton}. With this approach, we compute the exciton energies and wave functions in monolayer WSe$_2$ for exciton momenta $\bfQ$ on a regular Brillouin zone (BZ) grid (see Appendix~\ref{Appendix:computation}). 
The resulting exciton band structure is shown in Fig.~\ref{Fig:band}(a), where we highlight the energy of the two degenerate optically-active (or \lq\lq bright'') excitons at $\Gamma$, also known as A-excitons in 2D-TMDs, whose computed energy of $E_{0}\!=\!1.665$ eV agrees with experiments~\cite{mouri2014nonlinear}. These two A-excitons consist, respectively, of an electron-hole pair in the electronic K- or K$'$-valleys~\cite{wang2018colloquium}. 
\\
\indent
The exciton band structure has two nearly-degenerate minima associated with a spin-singlet dark exciton at $\rm Q$ and a spin-triplet dark exciton at $\rm M$, with the latter lower by \mbox{20 meV} due to the lack of exchange repulsion. 
In Fig.~\ref{Fig:band}(b), we show the main electronic transitions that make up the dark excitons with momenta $\bfQ= \rm K$, Q, and M, together with the corresponding electron-hole pairs in the electronic BZ. For these dark excitons, the hole occupies the valence band edge at the K or K$'$ BZ corner, and the electron occupies the conduction band minima at K or Q. 

\subsection{Exciton-phonon interactions and PL linewidth}
\vspace{-10pt}
\begin{figure}[!b]
\centering
\includegraphics[scale=0.70]{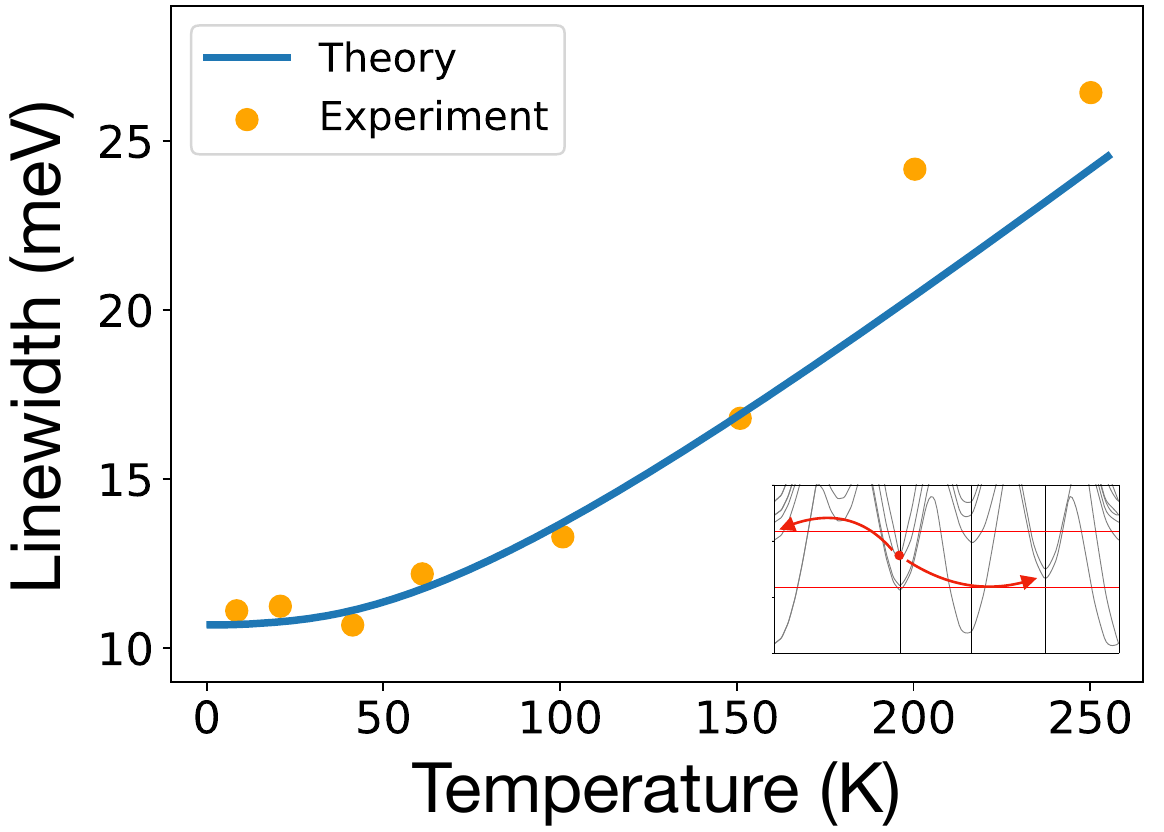}
\caption{Computed PL linewidth from ex-ph interactions in monolayer WSe$_2$, shown as a function of temperature and compared with experiment~\cite{chellappan2018effect}. The inset shows two main ex-ph scattering processes for bright exciton relaxation, $\Gamma$ to M and $\Gamma$ to K. The computed results are shifted upward by 7 meV to match experiment.} 
\label{Fig:width}
\end{figure}
We compute the ex-ph interactions in monolayer WSe$_2$ starting from the $e$-ph interactions and the BSE exciton wave functions, using an approach we developed in Ref.~\cite{chen2020exciton} (see Appendix~\ref{Appendix:RT}). The resulting ex-ph matrix elements $\mathcal{G}_{nm\nu}(\mathbf{Q} ,\mathbf{q} )$ describe the probability amplitude for an exciton in state $|S_n(\bfQ)\rangle$ to transition to state $|S_m(\bfQ+\bfq)\rangle$ when scattered by a phonon with mode $\nu$ and momentum $\bfq$. Our $e$-ph, BSE, and ex-ph calculations include spin-orbit coupling by using fully relativistic pseudopotentials. 
\\
\indent 
We validate the accuracy of our ex-ph interactions by computing the intrinsic PL linewidth due to ex-ph processes and comparing it with experiments. 
We focus on the bright A-exciton, which can recombine radiatively by emitting circularly polarized light~\cite{xiao2007valley}, and compute its ex-ph scattering rate to obtain the intrinsic PL linewidth~\cite{kira2006many,chen2020exciton}:
\begin{align}
\label{Eq:width}
\Gamma_{n\bfQ} (T)&=\frac{2\pi}{\mathcal{N}_\bfq}
\sum_{m\nu\bfq} |\mathcal{G}_{n,m\nu}(\bfQ,\bfq)|^2\nn\\
&\!\!\!\!\!\times
\left[(N_{\nu\bfq}+1+F_{m\bfQ+\bfq})
\times\d(E_{n \bfQ}-E'_{m\bfQ+\bfq}-\hbar\w_{\nu\bfq})\right.
\nn\\
&~~
\left.+
(N_{\nu\bfq}-F_{m\bfQ+\bfq})
\times\d(E_{n \bfQ}-E'_{m\bfQ+\bfq}+\hbar\w_{\nu\bfq})
\right],
\end{align}
where we set $n=3$ and $\bfQ=0$ for the A-exciton, and $\mathcal{N}_\bfq$ is the number of $\bfq$-points in the BZ, $E_{m\bfQ}$ are exciton energies, and $\omega_{\nu\bfq}$ are phonon frequencies; $N_{\nu \bfq}(T)$
is the thermal occupation for phonons and $F_{m\bfQ}(T)$ for excitons, both satisfying the Bose-Einstein distribution at temperature $T$.  
\\
\indent
The temperature dependence of our computed PL linewidth, shown in Fig.~\ref{Fig:width}, agrees with experiments from Ref.~\cite{chellappan2018effect}, although to match the experimental curve we need to apply a 7 meV rigid upward shift. 
We attribute this difference to temperature independent factors not considered here, such as broadening from radiative decay and interactions with substrate and defects~\cite{Ajayi2017}. Note that simply rescaling the ex-ph matrix elements would lead to changes in the temperature dependence of the PL linewidth, which is governed by the exciton and phonon dispersions and by the momentum dependence and magnitude of the ex-ph interactions. Therefore, the agreement with experiment shows that these quantities are properly described in our calculations.
In the inset of Fig.~\ref{Fig:width}, we highlight the two main scattering channels responsible for the PL linewidth, whereby the bright exciton at $\Gamma$ is scattered to the M-valley by phonon absorption or to the K-valley by phonon emission, consistent with recent experimental evidence~\cite{liu2015observation}. 
%
%% -----   REAL TIME DYNAMICS   ----- %%
%
\section{Results}
\subsection{Real-time exciton dynamics\label{Subsect:BTE}}
\begin{figure}[!t]
\centering
\includegraphics[scale=0.32]{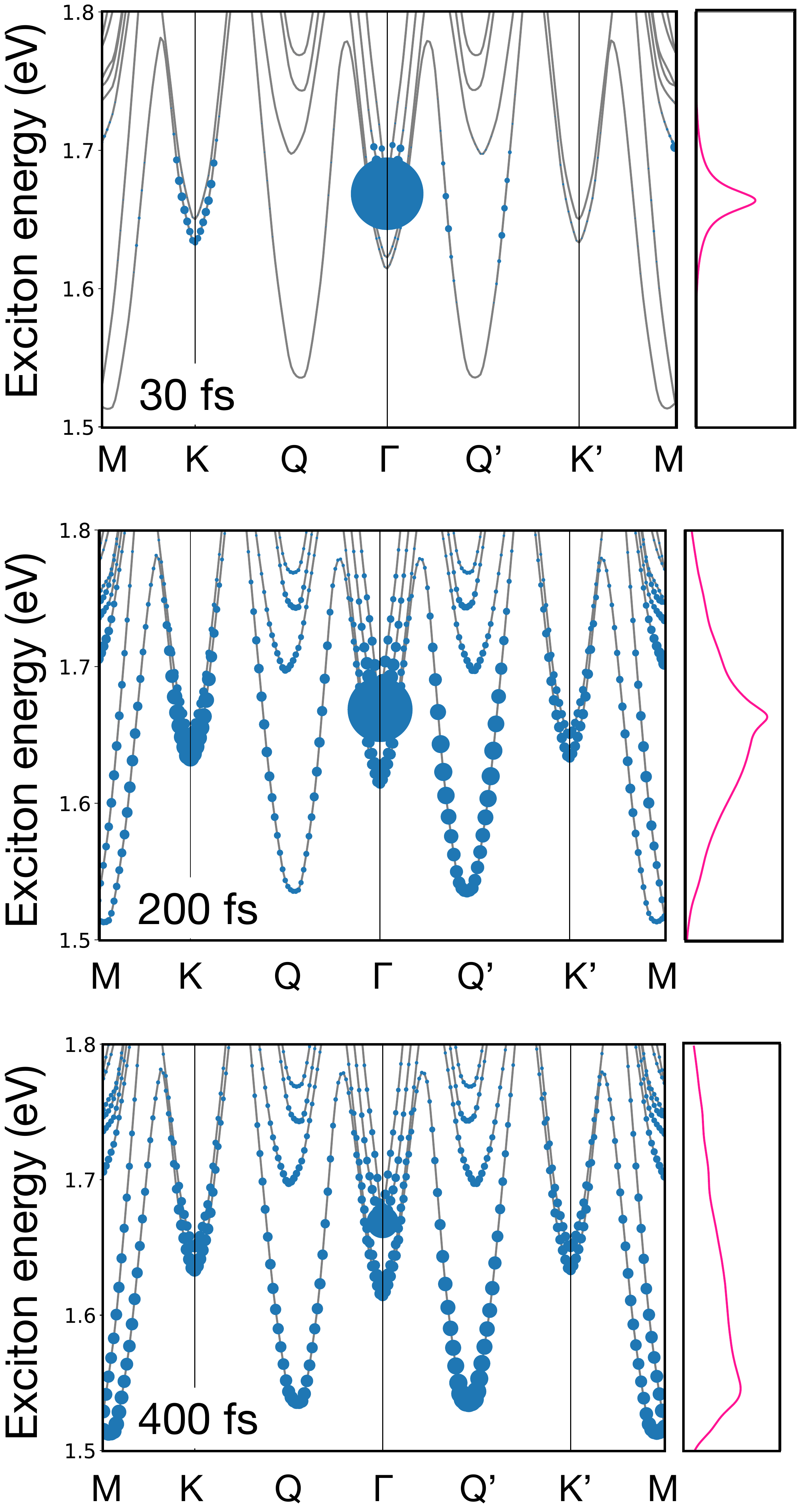} 
\caption{Snapshots of exciton relaxation in monolayer WSe$_2$ at 300~K, shown at times $t=$ 30, 200, and 400 fs from top to bottom.
We map the exciton populations onto the exciton band structure, with a spot radius proportional to the logarithm of the exciton populations, $\log(F_{n\bfQ})$. Next to each panel we give the exciton energy distribution (in arbitrary units, shown with magenta solid lines).  
}
\label{Fig:snapshot}
\end{figure}
\vspace{-10pt}
We simulate real-time exciton dynamics in monolayer WSe$_2$ using the ex-ph interactions discussed above. 
Analogous to the case of electron dynamics, 
we describe the exciton population dynamics using a real-time Boltzmann transport equation (rt-BTE)~\cite{jhalani2017ultrafast,tong2021toward, caruso2021nonequilibrium}. The key difference from the standard formalism for $e$-ph scattering is that excitons follow the Bose-Einstein statistics~\cite{Snoke}. Therefore, we derive a rt-BTE for exciton population dynamics, obtained by extending to bosonic populations the rt-BTE for electrons:
\begin{align}
\label{Eq:BTE-ex}
&\left(\frac{\partial F_{n\bfQ}}{\partial t}\right)^{\rm ex-ph}=
-\frac{2\pi}{\hbar}\frac{1}{\mathcal{N}_\bfq} \sum_{m\nu\bfq}
\left|\mathcal{G}_{nm\nu}(\bfQ,\bfq)\right|^2\nn\\
&~~~~~
\times\left[\delta\left(E_{n\bfQ}-E_{m\bfQ+\bfq}+\hbar\omega_{\nu\bfq}\right)\cdot F_{\rm abs}(t)\right.\nn\\
&~~~~~~~~+
\left.\delta\left(E_{n\bfQ}-E_{m\bfQ+\bfq}-\hbar\omega_{\nu\bfq}\right)\cdot 
F_{\rm em}(t)
\right]
\end{align}
where $F_{n\bfQ}(t)$ are time-dependent exciton populations. 
\\
\indent
The terms $F_{\rm abs} (t)$ and $F_{\rm em}(t)$ are occupation factors for phonon absorption and emission processes, defined by extending to excitons (treated as bosons) the corresponding equations for electrons~\cite{jhalani2017ultrafast,tong2021toward}:
\begin{subequations}
\label{Eq:FabsFem}
\begin{align}
&
F_{\rm abs} (t) = F_{n\bfQ}(t) N_{\nu\bfq}[1+F_{n\bfQ+\bfq}(t)]
\nn\\
&~~~~~~~~-[1+F_{n\bfQ}(t)](1+N_{\nu\bfq})F_{n\bfQ+\bfq}(t)\\
&
F_{\rm em} (t) = F_{n\bfQ}(t)(1+N_{\nu\bfq})[1+F_{n\bfQ+\bfq}(t)]
\nn\\
&~~~~~~~~-[1+F_{n\bfQ}(t)]N_{\nu\bfq}F_{n\bfQ+\bfq}(t),
\end{align}
\end{subequations}
where we assume constant phonon populations $N_{\nu\bfq}$ set to their thermal equilibrium value. 
\begin{figure*}[!t]
\centering
\includegraphics[scale=0.32]{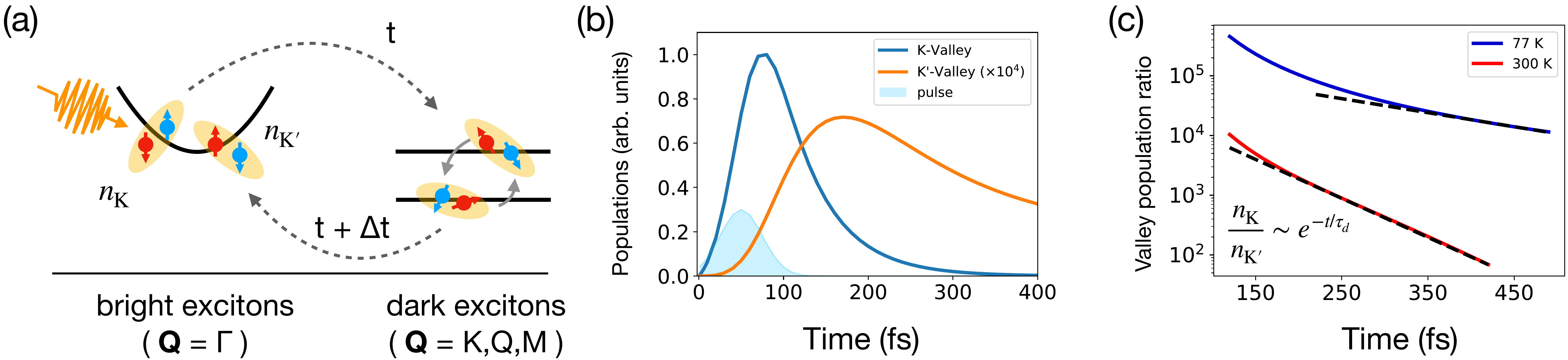}
\caption{(a) Schematic of exciton valley depolarization involving intermediate dark excitons. The bright A-excitons at $\Gamma$, corresponding to electron-hole pairs in the electronic K- and K$'$- valleys, can scatter into each other in a process mediated by dark excitons. (b) Exciton populations in the electronic K and K$'$ valleys following a simulated circularly-polarized 50 fs Gaussian pump pulse at 300~K temperature. 
(c) Corresponding ratio of the K and K$'$ valley populations after the pump pulse. The depolarization times extracted from the simulations are $\tau_{\rm d} =185$~fs at 77~K and $\tau_{\rm d} =66$~fs at 300~K.}
\label{Fig:valley occupation}
\end{figure*}
These occupation factors take into account the bosonic nature of excitons $-$ for example, the first term in Eq.~(\ref{Eq:FabsFem}a) implies that the exciton population is unchanged by phonon absorption if the initial exciton state or the specific phonon mode are unoccupied ($F_{n\bfQ}=0$ or $N_{\nu\bfq}=0$). 
Conversely, the first term in Eq.~(\ref{Eq:FabsFem}b) describes the exciton population change during phonon emission, for which a finite phonon occupation is not required. 
In both equations, the second terms account for the time reversal process of the respective first terms, while the factors $(1+F)$ reflect the exciton bosonic character and replace the electronic $(1-f)$ factors~\cite{jhalani2017ultrafast}, which are responsible for Pauli blocking~\cite{shah2013ultrafast}.
\\
\indent
The numerical calculations, implemented in the \mbox{{\sc Perturbo} code~\cite{zhou2020perturbo}}, time-step the exciton rt-BTE in Eq.~(\ref{Eq:BTE-ex}) with the Euler algorithm using a 1 fs time step. 
In all discussions below, we employ a right-handed circularly polarized Gaussian pump pulse with a 50 fs full-width at half maximum, and measure time from the beginning of the pump pulse, which is taken as time $t\!=\!0$. This pulse generates bright excitons at $\Gamma$ consisting of electron-hole pairs in the electronic K-valley. Throughout the simulations, the phonon populations are kept to their thermal equilibrium value at the chosen temperature, a valid assumption in the low-excitation limit~\cite{shah2013ultrafast}. 
Additional details are provided in Appendix~\ref{Appendix:RT}. 
\\
\indent 
Snapshots of the exciton populations at times $t=$ 30, 200, and 400 fs are shown in Fig.~\ref{Fig:snapshot} for a simulation at 300~K temperature. At time $t=30$~fs, when the circularly-polarized pump pulse is still on, the excitons reside primarily in the lowest bright-exciton state at $\Gamma$, and a small fraction of excitons have transferred to the excitonic K-valley. 
An even smaller fraction of excitons absorb a phonon and scatter into exciton states at M with energy $E=1.7$~eV, or emit a phonon and transition to the Q$'$-valley. In the $t=200$ fs panel in Fig.~\ref{Fig:snapshot}, after the pump has been turned off and excitons have relaxed more extensively, the effects of valley polarization become apparent. Our initial excitation of the \textit{electronic} K-valley results in a greater occupation of the \textit{excitonic} K- and Q$'$-valleys compared to the K$'$- and Q-valleys; the opposite trend is observed for a pump pulse with opposite-handed circular polarization. 
\\
\indent
Finally, at time $t=400$~fs, most of the excitons have relaxed to the global minima, as shown by the exciton energy distributions in Fig.~\ref{Fig:snapshot}, and the exciton populations at K and $\Gamma$ are significantly reduced compared to their values at 200 fs, while the Q$'$- and M-valleys have the highest occupations. These results allow us to determine the bright A-exciton relaxation time in monolayer WSe$_2$, with a value of $\sim$0.5~ps at 300~K based on our simulation. 
\\
\subsection{Exciton valley depolarization}
\vspace{-10pt}
In monolayer TMDs, excitons can be generated selectively in the electronic K- or K$'$-valleys, using respectively right- or left-handed circularly polarized light~\cite{xiao2007valley}. An exciton generated in the K-valley will ultimately scatter into the K$'$-valley; this process, known as exciton valley depolarization, can be studied using pump-probe spectroscopy with circularly polarized light~\cite{wang2018colloquium}. 
Because of spin-valley locking~\cite{xiao2007valley}, intervalley scattering from $e$-ph interactions is relatively weak in 2D-TMDs, especially for hole carriers. 
Consequently, the role of phonons in exciton valley depolarization has been largely ignored in previous work, with the exception of a recent study showing a significant contribution to intervalley spin-flip scattering from $e$-ph interactions~\cite{wang2018intravalley}. 
When excitons are taken into account, intervalley spin-flip processes can be mediated by dark excitons without a well-defined spin character, opening up a broader phase space for intervalley scattering, as shown pictorially in Fig.~\ref{Fig:valley occupation}(a). 
\\
\indent
Here we examine in detail these bright-to-dark exciton scattering processes. Using exciton rt-BTE simulations at two temperatures (77~K and 300~K), we compute phonon-induced exciton valley depolarization times using the time-dependent exciton populations in the electronic K- and K$'$-valleys. Figure~\ref{Fig:valley occupation}(b) shows results at 300~K, where the K-valley exciton population increases rapidly during the circularly-polarized pump pulse (first $\sim$100~fs) and then decreases due to ex-ph scattering. 
Meanwhile, the ${\rm K}'$-valley exciton population increases and reaches a maximum $\sim$100~fs after the K-valley exciton population has peaked, after which it decreases slowly, reaching about half of its peak value at 400~fs. 
\\
\indent
We quantify the valley depolarization time by fitting the ratio of the A-exciton populations in the electronic K- and K$'$-valleys, which correspond to the BSE states with $n\!=\!(3,4)$ and momentum $\bfQ\!=\!0$. Defining as $n_{\rm K} \!\equiv\! F_{n=3,\bfQ=0}$ the exciton population in the electronic K-valley, and $n_{{\rm K}'}\! \equiv\! F_{n=4,\bfQ=0}$ the exciton population in the electronic K$'$-valley, we write the K/K$'$-valley exciton population ratio as
\begin{equation}
\frac{n_{\rm K}}{n_{{\rm K}'}}=A\, e^{-(t-t_0)/\tau_{\rm d}},
\end{equation}
where $A$ and $t_0$ are constants dependent on simulation settings, such as pulse strength and duration, and the time constant $\tau_{\rm d}$ is the phonon-induced exciton valley depolarization time. 
In Fig.~\ref{Fig:valley occupation}(c), by fitting the ratio $n_{\rm K}/{n_{\rm K'}}$ between 170$-$420~fs, we obtain exciton depolarization times for monolayer WSe$_2$ of $\tau_d=185$~fs at 77~K and $\tau_d=66$~fs at 300~K. Beyond $\sim$500~fs excitons occupy equally the electronic K and K$'$ valleys, and the initial valley polarization is entirely lost. 
Our computed exciton valley depolarization time of $185$~fs at 77~K is in agreement with recent experiments by Wang \textit{et al.}~\cite{wang2018intravalley}, who measured an intervalley depolarization time of $\sim$500~fs at 77~K~(see Fig. 6(c) in Ref.~\cite{wang2018intravalley}). Note that their measurements were carried out in monolayer WS$_2$, whose exciton physics is closely related to monolayer WSe$_2$ studied here. Contributions from exciton-exciton scattering, not considered in this work, may further improve the agreement with experiment.
\\
\indent
We compare these results with the single-particle picture, where exciton valley depolarization is limited by intervalley scattering of the hole carrier, which is slower than electron intervalley scattering due to the greater spin-orbit coupling effect in the valence band~\cite{xiao2007valley}. 
Recent work has predicted the timescale for spin-flip intervalley scattering of hole carriers in monolayer WSe$_2$, with computed values of about 3 ns at 77~K and \mbox{0.1 ns} at 300~K~(see Fig. 2(c) in Ref.~\cite{parkPredicting2022}), using a first-principles method that can accurately predict spin relaxation times~\cite{parkPredicting2022}. 
These long timescales for hole intervalley scattering, which are a bottleneck for exciton valley depolarization in the single-particle picture, are 3$-$4 orders of magnitude greater than those computed here by properly including excitonic effects and dark-exciton scattering. This comparison with the single-particle picture demonstrates the key role played by the dark excitons, which can mediate exciton valley depolarization and speed it up by orders of magnitude.
It additionally emphasized the shortcomings of analyzing excitonic processes based on the dynamics of independent electron and hole carriers.

\section{Discussion}
\vspace{-10pt}
The main exciton relaxation pathways revealed by our first-principles calculations give rise to distinctive excitonic signatures in time-domain spectra. Here we discuss this point and showcase the ability of our approach to predict excitonic effects in time-domain spectroscopies. 
%
% SIMULATED TR-ARPES
%
\subsection{Simulated time-resolved ARPES}
\vspace{-10pt}
Recent experiments have shown that tr-ARPES is a valuable tool to study exciton dynamics because it can probe the time evolution of electron and hole carriers making up the bound  excitons~\cite{Gedik,Dani1,Dani2}.
\begin{figure}[!t]
    \centering
    \includegraphics[scale=0.35]{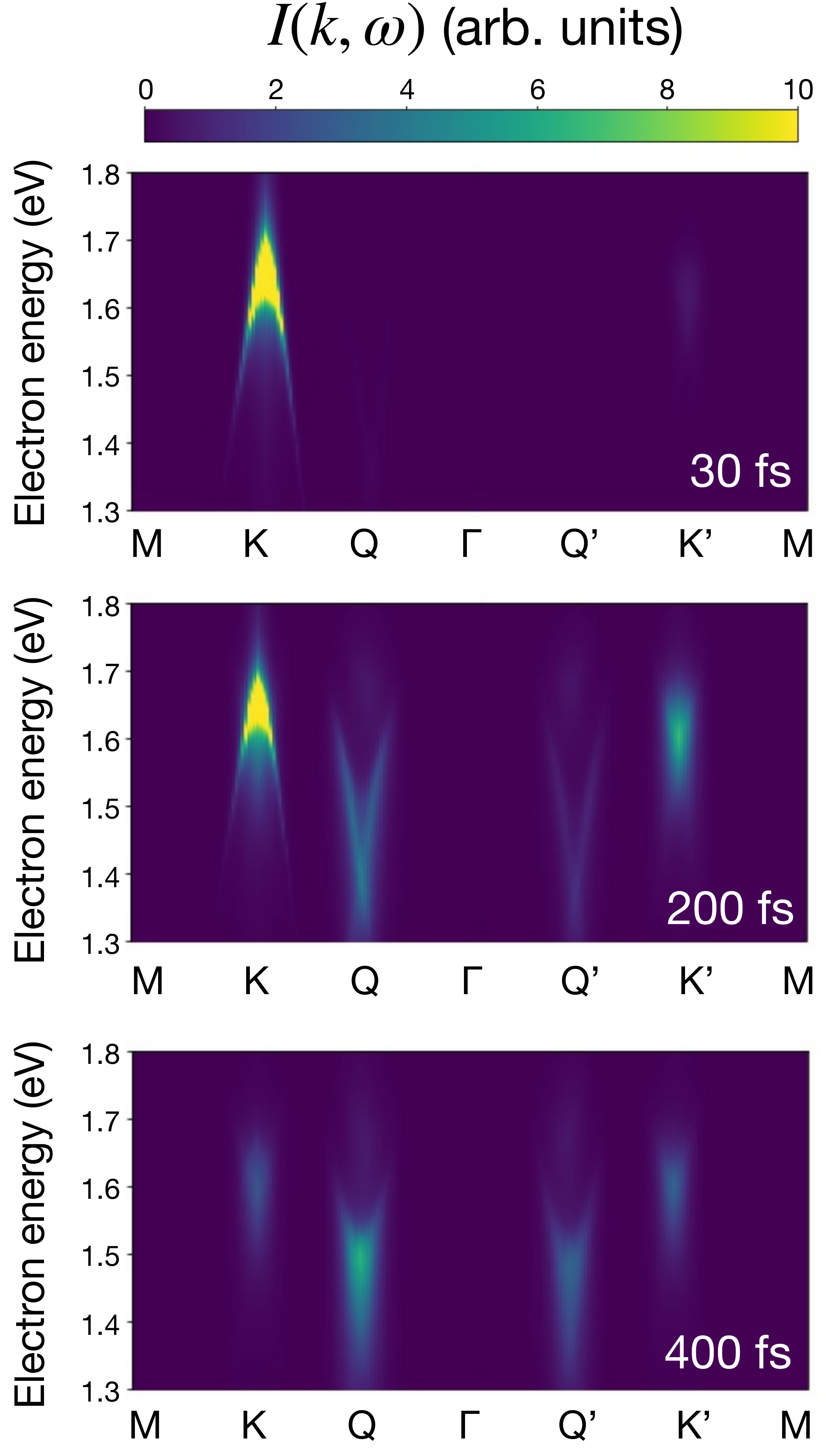}
    \caption{Simulated tr-ARPES spectra at times $t=30, 200$ and 400 fs, colored according to the computed tr-ARPES signal $I(k,\omega; t)$ in Eq.~(\ref{Eq:ARPES}) and plotted as a function of electron crystal momentum.}
    \label{fig:my_arpes}
\end{figure}
Similar to standard ARPES technique, tr-ARPES measures the energy and momentum distribution of photoemitted electrons, but it can also characterize their nonequilibrium dynamics with sub-ps time resolution~\cite{Gedik}. 
The tr-ARPES signal $I(\bfk,\w;t)$ is the product of electron occupations at time $t$ and the corresponding spectral functions. When the photoelectrons are emitted from bound excitons, one can relate the tr-ARPES signal to the exciton populations using~\cite{perfetto2016first,sangalli2021excitons}: 
\begin{align}
\label{Eq:ARPES}
    &I(k,\w;t)\propto \nn\\
    &~~\textrm{Im}
    \Bigg(
    \sum_{m\bfQ}F_{m\bfQ}(t)
    \sum_{cv}\frac{|A^{m\bfQ}_{vc\bfk}|^2}{\w-(E_{m\bfQ}+\e_{v\bfQ-\bfk})+i\eta}
    \Bigg),
\end{align}
where $F_{m\bfQ}(t)$ are time-dependent exciton populations from the rt-BTE,
and the square exciton wave functions $|A_{cv\bfk}^{m\bfQ}|^2$ are the probability to find an electron, bound in the exciton state $(m,\bfQ)$, at energy $E_{m\bfQ}+\e_{v\bfQ-\bfk}$, where $\e_{v\bfQ-\bfk}$ is the hole energy (see Eq.~(\ref{Eq:BSE-EA})); the broadening $\eta$ is set to the exciton linewidth in Eq.~(\ref{Eq:width}). 
Based on this expression, the tr-ARPES signal is expected to exhibit a copy of the valence band shifted by the exciton energy-momentum dispersion $E_{m\bfQ}$ and weighted by the exciton wave function.
\\
\indent
The simulated tr-ARPES spectra, obtained by converting the rt-BTE exciton populations to photoemission signal via Eq.~(\ref{Eq:ARPES}), are shown in Fig.~\ref{fig:my_arpes} at three different times. During the right-handed circularly polarized pump pulse, electrons making up the bright A-excitons occupy a small region in the electronic K-valley. Similar to a previous model of tr-ARPES~\cite{christiansen2019theory}, at early time during the pump pulse ($t=30$~fs) our simulated tr-ARPES signal shows a downward parabola resembling the electronic valence band dispersion. This trend is a signature of optically pumped excitons with vanishing momentum dominating the exciton populations. 
At time $t=200$~fs, excitons with finite momentum give a photoemitted electron signal at K$'$ with energy lower by 10$-$50 meV than the bright-exciton signal at K. We additionally find electrons in the Q-valley deriving from Q$'$-excitons. 
Finally, at 400 fs the signal at K and K$'$ is weak, and the signal at Q reveals an upward parabolic band with a broad minimum due to overlapping contributions from the M, Q, and Q$'$ excitons. These results show that the exciton populations from the rt-BTE can be employed directly to predict tr-ARPES spectra and aid microscopic interpretation of the underlying exciton dynamics.\\

\subsection{Transient absorption}
\vspace{-10pt}
The redistribution of valence electrons modifies the optical response of a material. Following photoexcitation of electrons in the conduction band and holes in the valence band, Pauli blocking slows down optical transitions, thus reducing the absorption coefficient~\cite{shah2013ultrafast}. 
Transient absorption in the presence of \mbox{excitons} has been studied primarily with models using simple approximations for exciton wave functions and transition dipoles~\cite{shinada1966interband,Schmitt1985theory,huang1990carrier}. 
Here we combine exciton data from the \textit{ab initio} BSE with time-dependent exciton populations from our rt-BTE, with the goal of achieving quantitative predictions of transient absorption spectra in the presence of excitons. We derive an expression for transient absorption accounting for excitonic effects (see Appendix~\ref{Appendix:transient}):
\begin{align}
\label{Eq:trans-abs}
&\Delta\a(\w,t)=
-\frac{\left|{\bfp}_n\cdot \hat{\bfe}\right|^2}
{\w-E_n+i\Gamma_n}\nn\\
&\times
{\rm Re}\left(\frac{\sum_{vc\bfk} (f_{c\bfk}(t)+f_{v\bfk}(t))(A^{S_n}_{vc\bfk}\bfp_{vc\bfk})\cdot \hat{\bfe}}
{{\bfp}_n\cdot \hat{\bfe}}\right),
\end{align}
where $\bfp_n=\<G|\bfp|S_n\>$ is the exciton transition dipole of the $n$-th excitonic state in the light-cone, $\bfp_{vc\bfk}=\<c\bfk|\bfp|v\bfk\>$ are interband electronic transition dipoles, and $\hat{\bfe}$ is the polarization of the probe. The populations of carriers belonging to bound excitons, denoted as $f_{c \bfk}(t)$ for electrons and $f_{v \bfk}(t)$ for holes, can be obtained from the exciton populations and the BSE exciton wave functions using:
\begin{align}
\label{eq:carriers}
&f_{c \bfk}(t)=\sum_{m\bfQ,v} F_{m\bfQ}(t)|A^{S_m(\bfQ)}_{vc\bfk}|^2
\nn\\
&f_{v \bfk}(t)=\sum_{m\bfQ,c} F_{m\bfQ}(t)|A^{S_m(\bfQ)}_{vc\bfk+\bfQ}|^2. 
\end{align}
\begin{figure}[!t]
%
%%%  FIGURE 4 --- TRANSIENT ABSORPTION
%
\includegraphics[scale=0.25]{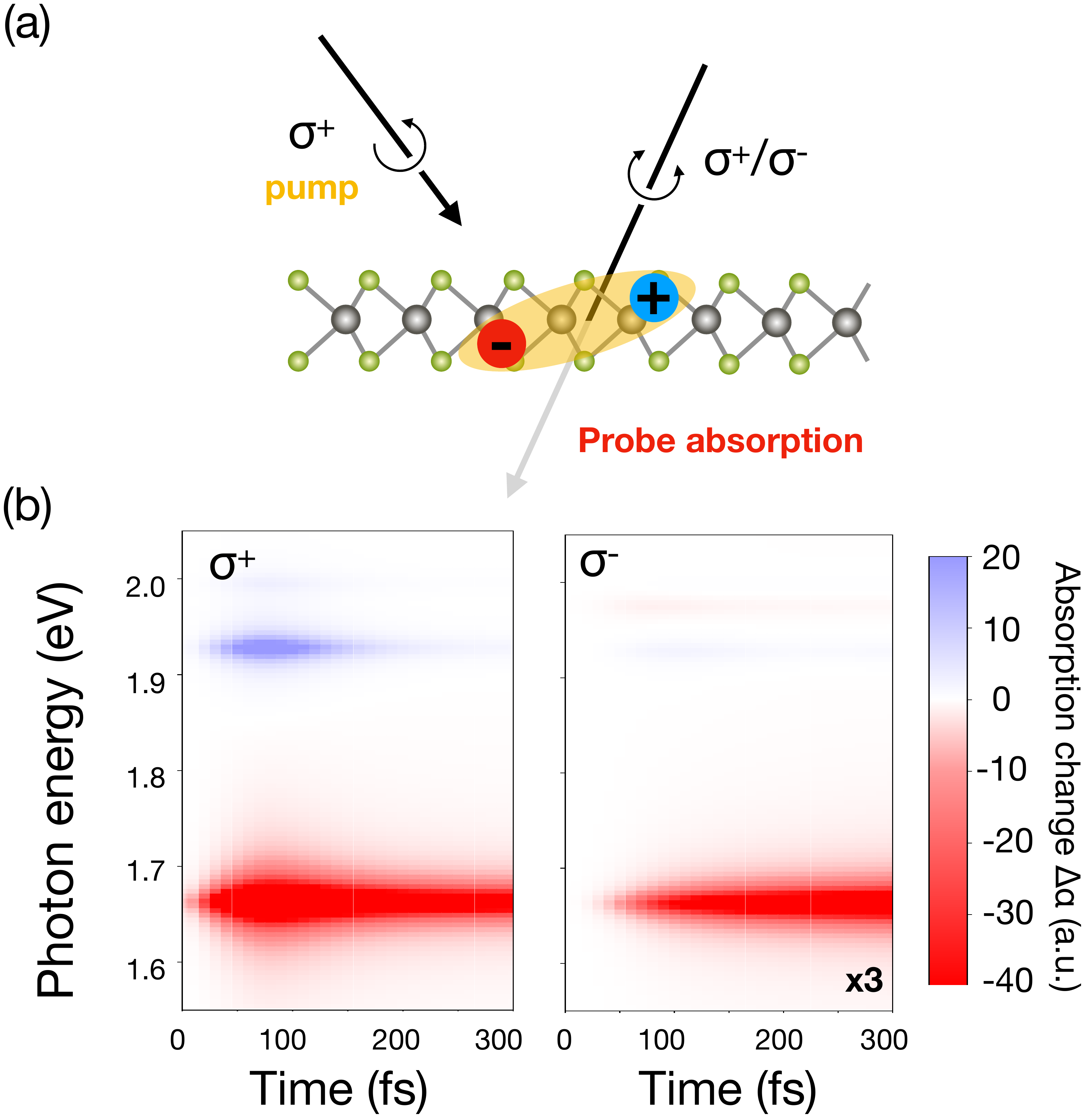}
\caption{(a) Schematic of our simulated pump-probe transient absorption settings. (b) Transient absorption spectra of monolayer $\rm WSe_2$ pumped with a right-handed circularly polarized ($\sigma^{+}$) incident pulse. We plot separately the transient absorption spectra up to 300 fs for probes with $\sigma^{+}$ (left panel) and $\sigma^{-}$ (right panel) polarizations.
}
\label{Fig:transient}
\end{figure}
\indent
To simulate transient absorption, we combine Eqs.~(\ref{Eq:trans-abs})-(\ref{eq:carriers}) with our rt-BTE exciton populations, and compute changes in the absorption coefficient for right- ($\s^+$) and left-handed ($\s^-$) circularly-polarized probes following a $\s^+$ pump pulse with 50 fs half-width, as shown schematically in Fig.~\ref{Fig:transient}(a). The simulated transient absorption spectra, shown for the first 300 fs in Fig.~\ref{Fig:transient}(b), contain features associated with exciton relaxation and \mbox{intervalley} scattering. 
\\
\indent
For the $\s^+$ probe, absorption of $\sigma^{+}$ light is reduced near the bright-exciton resonance at $\sim$1.665~eV due to Pauli blocking; after the pump pulse is turned off, starting at 75 fs the absorption gradually increases toward the equilibrium value as the A-excitons relax via ex-ph scattering. 
Different from the single-particle picture, where transient absorption due to Pauli blocking is always negative, the excitonic transient absorption in Eq.~(\ref{Eq:trans-abs}) can be either positive or negative.
For example, at higher energies of 1.93 and 1.98 eV, we find positive transient absorption values up to 200 fs due the enhancement of the effective transition dipoles in the presence of finite exciton populations (see Appendix~\ref{Appendix:transient}).
\\
\indent
Conversely, for the $\s^-$ probe, Pauli blocking plays only a minor role as the absorption of $\s^-$ polarized light is in general weak when the pump pulse is $\s^+$ polarized. In this case, we observe only negative transient absorption, with a 50 fs time delay after the $\s^+$ pump is turned on and increasing monotonically in time during our simulation due to the increasing exciton population in the electronic K$'$-valley. These results show that our formalism can capture excitonic signatures in transient absorption spectra. 

\section{Conclusion}
\vspace{-10pt}
We presented an approach to study exciton dynamics using first-principles ex-ph interactions. Our method is based on a bosonic Boltzmann equation combined with material-specific exciton properties from the \textit{ab initio} BSE, thus going beyond heuristic models of exciton dynamics.  
Our calculations on monolayer $\rm WSe_2$ predict sub-ps timescales for exciton valley depolarization, in contrast with the ns timescale for this process in the single-particle picture. These results highlight the need for methods addressing explicitly exciton interactions and dynamics $-$ for both bright and dark excitons $-$ as we do in this work. We also show simulations of ultrafast exciton spectroscopies, including tr-ARPES and transient absorption, in both cases going beyond the single-particle picture and capturing excitonic effects. 
Future work will explore exciton-induced lattice dynamics using this formalism. Taken together, this work advances first-principles calculations of exciton dynamics and time-domain spectroscopies, paving the way for quantitative studies of nonequilibrium physics in materials with strongly bound excitons. 

\begin{acknowledgments}
The authors thank Davide Sangalli, Ivan Maliyov and Jinsoo Park for fruitful discussions. 
This material is based upon work supported by the U.S. Department of Energy, Office of Science, Office of Advanced Scientific Computing Research and Office of Basic Energy Sciences, Scientific Discovery through Advanced Computing (SciDAC) program under Award Number DE‐SC0022088, which supported method development. M.B. was partially supported by the Liquid Sunlight Alliance, which is supported by the U.S. Department of Energy, Office of Science, Office of Basic Energy Sciences, under Award No. DE-SC0021266. Code development was partially funded by the National Science Foundation under Grant No.~OAC-2209262. H.-Y. Chen was partially supported by the J.~Yang Fellowship.  This research used resources of the National Energy Research Scientific Computing Center (NERSC), a U.S. Department of Energy Office of Science User Facility located at Lawrence Berkeley National Laboratory, operated under Contract No.~DE-AC02-05CH11231. 
\end{acknowledgments}

\newpage
\appendix
\section{Computational details \label{Appendix:computation}}
The DFT calculations on monolayer $\rm WSe_2$ are carried out using a relaxed lattice constant of 3.27~\AA ~and a 20~\AA ~layer-normal separation between periodic replicas. The ground state is computed using DFT in the generalized gradient approximation~\cite{GGA} with the {\sc{Quantum ESPRESSO}} code~\cite{giannozzi2009quantum}. We employ a 60 Ry kinetic energy cutoff to compute the electronic structure on a $72 \times 72 \times 1$ uniform $\bfk$-point grid in the BZ. The lattice dynamics and $e$-ph perturbation potentials are computed with density functional perturbation theory~\cite{baroni2001phonons} on a $ 36 \times 36 \times 1$  $\bfq$-point grid using {\sc{Quantum ESPRESSO}}. We use the {\sc Perturbo} code \cite{zhou2020perturbo} to compute the $e$-ph matrix elements on these electron and phonon momentum grids. All calculations include spin-orbit coupling by using fully relativistic norm-conserving pseudopotentials~\cite{Spaldin-PP} generated with Pseudo Dojo~\cite{van2018pseudodojo}. 
\\
\indent
To study excitons, we carry out first-principles BSE calculations with the {\sc Yambo} code~\cite{sangalli2019many}. We solve the finite-momentum BSE exciton Hamiltonian~\cite{Qiu,Cudazzo}
\begin{align}
     \label{Eq:BSE}
H^{(\bfQ)}_{vc\bfk,v'c'\bfk'}=&\< vc\bfk|H^{(\bfQ)}|v'c'\bfk'\>\nn\\
=&(\e_{c\bfk}-\e_{v\bfk-\bfQ})\delta_{vv'} \delta_{cc'}\delta_{\bfk\bfk'}+K^{(\bfQ)}_{vc\bfk,v'c'\bfk'}\nn\\
\end{align}
where the kernel $K^{(\bfQ)}_{vc\bfk,v'c'\bfk'}$ describes the electron-hole interactions for excitons with momentum $\bfQ$. The BSE solution provides exciton energies $E_{S(\bfQ)}$ and wave functions $|S(\bfQ)\>$ expressed in the transition basis, with expansion coefficients $A_{vc\bfk}^{S(\bfQ)}$ satisfying~\cite{Qiu,Cudazzo}
\begin{eqnarray}
\label{Eq:BSE-EA}
&&\sum_{c'v'\bfk'}H^{(\bfQ)}_{vc\bfk,v'c'\bfk'}A_{v'c'\bfk'}^{S(\bfQ)}=E_{S(\bfQ)}A_{vc\bfk}^{S(\bfQ)}\nn\\
&&~~~~~~~~|S(\bfQ)\>=\sum_{vc\bfk}A_{vc\bfk}^{S(\bfQ)}|c\bfk\>|v\bfk-\bfQ\>,
\end{eqnarray}
where $v$ and $c$ are valence and conduction band indices, and $\e_{v\bfk-\bfQ}$ and $\e_{c\bfk}$ are the corresponding quasiparticle energies. 
In these calculations, we compute the screened Coulomb interaction using a 5 Ry cutoff and 300 bands, and use the 2 highest valence bands and 2 lowest conduction bands to compute the BSE kernel; the quasiparticle energies are fine-tuned to reproduce the experimental band gap and the relative valley energies~\cite{zhang2015probing}. 
The BSE is solved on a $ 36 \times 36 \times 1$ exciton momentum $\bfQ$-grid.

\section{Exciton-phonon interactions and exciton dynamics\label{Appendix:RT}}
\vspace{-10pt}
We implement the exciton rt-BTE in a developer version of the {\sc Perturbo} code~\cite{zhou2020perturbo}. 
The ex-ph matrix elements $\mathcal{G}_{nm\nu}(\mathbf{Q} ,\mathbf{q} )$ used in the rt-BTE, which quantify the coupling between excitons and phonons as discussed above, are computed by combining first-principles BSE and $e$-ph calculations using~\cite{chen2020exciton}:
\begin{align}
\label{Eq:exph_coupling}
&\mathcal{G}_{nm\nu}(\bfQ,\bfq)~\nn\\
&~=\sum_{\bfk}\left[
\sum_{vcc'}
A^{S_m(\bfQ+\bfq)*}
_{vc(\bfk+\bfq)}
A^{S_n(\bfQ)}_{vc'\bfk}
g_{c'c\nu}(\bfk,\bfq)\right.~\nn\\
&~~~
\left.-
\sum_{cvv'}
A^{S_m(\bfQ+\bfq)*}_{vc\bfk}
A^{S_n(\bfQ)}_{v'c\bfk}
g_{vv'\nu}(\bfk-\bfQ-\bfq,\bfq)
\right],
\end{align}
where $g_{c'c\nu}(\bfk,\bfq)$ and $g_{vv'\nu}(\bfk,\bfq)$ are $e$-ph matrix elements~\cite{zhou2020perturbo}. 
In this work, the ex-ph matrix elements are computed directly on the same $ 36 \times 36 \times 1$ uniform grid used for phonons and excitons. For the rt-BTE and PL calculations, we use linear interpolation to obtain the ex-ph matrix elements and exciton energies on finer grids with $216~\times~216~\times~1$ $\bfq$- and $\bfQ$-points. The real-time simulations are performed on this grid using the 10 lowest exciton bands. 
\\
\indent
In our real-time simulations, coupling of the exciton-phonon system with the photon field is included although it is not shown explicitly in the exciton rt-BTE in Eq.~(\ref{Eq:BTE-ex}). To account for the circularly-polarized pump pulse, we increase the $n\!=\!3$ A-exciton population proportionally to the pump intensity during the pump pulse. This approach neglects short-lived interband coherences and focuses on the generation of exciton populations in the first $\sim$100 fs of the simulation (i.e., while the pump is on), consistent with the population-based picture of the rt-BTE. In addition, we approximately account for radiative recombination by decreasing the populations of both A-exciton states with a time constant of $\tau_{\rm rad}=0.22$~ps equal to the intrinsic A-exciton radiative lifetime in monolayer WSe$_2$~\cite{palummo2015exciton}.

\section{Excitonic transient absorption \label{Appendix:transient}}
\vspace{-10pt}
We derive the transient absorption formula in Eq.~(\ref{Eq:trans-abs}) by extending the model in Ref.~\cite{huang1990carrier} to the $ab~initio$ BSE formalism. In the presence of photoexcited carriers in bound excitons, the exciton wave function follows the transition-basis expansion under the modulation of Pauli exclusion:
\begin{equation}
|S_n\>=\sum_{vc\bfk} A^{S_n}_{vc\bfk}(1-f_{c\bfk}-f_{v\bfk})^{1/2}|\bfk,c\>|\bfk,v\>
\end{equation}
where $f_{c\bfk}(t)$ are electron and $f_{v\bfk}(t)$ hole populations for carriers belonging to bound excitons, which we compute using Eq.~(\ref{eq:carriers}) from the time-dependent exciton populations. The corresponding exciton transition dipoles become
\begin{equation}
\tilde{\bfp}_n=\sum_{vc\bfk}
A^{S_n}_{vc\bfk}(1-f_{c\bfk}-f_{v\bfk})^{1/2}\bfp_{vc\bfk},
\end{equation}
where $\tilde{\bfp}_n$ denotes the dipole modulated by the carriers in bound excitons (as opposed to the intrinsic dipole, $\bfp_n$).
\\
\indent
For incident light with frequency $\w$ and polarization $\hat{\bfe}$, we define the change in optical absorption due to the carriers in bound excitons as
\begin{equation}
\Delta \a(\w)=\sum_n\frac{|(\tilde{\bfp}_n-\bfp_n)\cdot \hat{\bfe}|^2}{\w-E_n+i\Gamma_n},
\end{equation}
where $n$ indexes the exciton states with $\bfQ=0$, which have energy $E_n$ and linewidth $\Gamma_n$. 
Expanding terms proportional to the electron and hole populations to lowest order, we get
\vspace{-6pt}
\begin{eqnarray}
\label{Eq:DIP}
&& |\tilde{\bfp}_n\cdot \hat{\bfe}|^2=\left|  \sum_{vc\bfk}
A^{S_n}_{vc\bfk}(1-f_{c\bfk}-f_{v\bfk})^{1/2}\bfp_{vc\bfk} \cdot \hat{\bfe}\right|^2\nn\\
&&
\approx
\left|  \sum_{vc\bfk}
A^{S_n}_{vc\bfk}\bfp_{vc\bfk} \cdot \hat{\bfe}-\sum_{vc\bfk}(\frac{f_{c\bfk}+f_{v\bfk}}{2})A^{S_n}_{vc\bfk}\bfp_{vc\bfk} \cdot \hat{\bfe} \right|^2
\nn
\\
&&
=
|{\bfp}_n\cdot \hat{\bfe}|^2
\left|  1-\left(\frac{\sum_{vc\bfk} (f_{c\bfk}+f_{v\bfk})(A^{S_n}_{vc\bfk}
\bfp_{vc\bfk}) \cdot \hat{\bfe}}
{2~{\bfp}_n\cdot \hat{\bfe}}\right) \right|^2
\nn
\\
&&
\approx 
|{\bfp}_n\cdot \hat{\bfe}|^2
\left[  1-{\rm Re}\left(
\frac{\sum_{vc\bfk} (f_{c\bfk}+f_{v\bfk})(A^{S_n}_{vc\bfk}
\bfp_{vc\bfk}) \cdot \hat{\bfe}}
{{\bfp}_n\cdot \hat{\bfe}}\right) \right].
\nn
\\
\end{eqnarray}
\vspace{-6pt}
Thus we obtain the final result used in Eq.~(\ref{Eq:trans-abs}):
\begin{align}
&\Delta\a(\w)= 
-\sum_{n}
\frac{|{\bfp}_n\cdot \hat{\bfe}|^2}
{\w-E_n+i\Gamma_n}
\nn\\&~~~
\times
{\rm Re}
\left(\frac{\sum_{vc\bfk} (f_{c\bfk}+f_{v\bfk})(A^{S_n}_{vc\bfk}\bfp_{vc\bfk}) \cdot \hat{\bfe}}
{{\bfp}_n\cdot \hat{\bfe}}\right).
\end{align}
In the absence of excitonic effects, this result reduces to the well-known transient absorption formula for independent electron and hole carriers.

%\newpage
\vspace{-6pt}
\bibliographystyle{apsrev4-2}
\bibliography{Ultra_WSe2_ref}
\end{document}